\documentstyle[twocolumn,aps,epsf]{revtex}
\tighten
\begin{document}
\draft
\title{Phase Diagram of the Extended Hubbard Model with Correlated-Hopping 
Interaction}

\author{G.I. Japaridze$^1$ and A.P. Kampf}

\address{
Institut f\"ur Physik, Theoretische Physik III, Elektronische Korrelationen
und Magnetismus,\\ 
Universit\"at Augsburg, 86135 Augsburg, Germany}

\address{~
\parbox{14cm}{\rm 
\medskip
A one-dimensional model of interacting electrons with on-site $U$, 
nearest-neighbor $V$, and correlated-hopping interaction $T^{\ast}$ is studied 
at half-filling using the continuum-limit field theory approach. The 
ground-state phase diagram is obtained for a wide range of coupling constants. 
In addition to the insulating spin- and charge-density wave phases for large 
$U$ and $V$, respectively, we identify bond-located ordered phases 
corresponding to an enhanced Peierls instability in the system for 
$T^{\ast}>0$, $|U-2V|<8T^{\ast}/\pi$ and to a staggered magnetization located 
on bonds between sites for $T^{\ast}<0$, $|U-2V|<8|T^{\ast}|/\pi$. The general 
ground state phase diagram including insulating, metallic, and superconducting 
phases is discussed.
 }}
\maketitle
\narrowtext

\section{\bf Introduction}

Since the discovery of high-$T_{c}$ superconductivity there is continuous 
interest in models of interacting electrons with unconventional correlation 
mechanisms. Among others, models with correlated-hopping (CH) interaction 
[1-37] are the subject of current studies. In addition to the usual 
interaction between electrons on the same site $(U)$ and/or on 
nearest-neigbor (nn) sites $(V)$, these models contain terms describing the 
modification of the electronic hopping motion by the presence of other 
particles. Such a term emerges rather naturally in the construction of a 
tight-binding Hamiltonian \cite{Hub} and describes the interaction between 
charges located on bonds and on lattice sites (the bond-charge interaction). 
Generally, a model with CH interaction can naturally be viewed 
either as an effective model obtained after integrating out additional degrees 
of freedom \cite{FF,SA,SAMP} or as a phenomenological model.

The CH model was first proposed by Foglio and Falikov in 1979 to describe the 
low-energy properties of mixed valence systems \cite{FF}. In the eighties the 
bond-charge coupling was discussed mainly in the context of organic conductors,
e.g. doped polyacetelene, to describe the interplay between Coulomb repulsion 
and Peierls dimerization effects \cite{KSSH,WSN,BHM,GC,Voit1,CGL}. 

The interest in models with CH interaction increased after 
the discovery of high-$T_{c}$ superconductivity. Hirsch was the first who 
pointed out that the CH interaction provides a mechanism for 
a superconducting instability \cite{Hirsch}. Soon 
after, E{\ss}ler, Korepin, and Schoutens proposed the integrable 
supersymmetric extension of the Hubbard model with a particular, strongly 
correlated "kinematics" and a truely superconducting ground state of the 
$\eta$-pairing type \cite{EKS}. These results were intensively explored later 
in the context of superconductivity in high-$T_{c}$ oxides: electrons with 
CH interaction were studied using the BCS type mean-field 
approach \cite{MH1,MH2}, the field theory renormalization-group treatment 
\cite{JM}, the exact solution for particular values of coupling constants, 
and by numerical techniques \cite{ES1,ES2,ES3,ES4,ES5,ES6}. Several exactly 
solvable 1D models of interacting electrons with CH coupling 
were proposed and intensively studied 
\cite{Refes1,Refes2,Refes3,Refes4,Refes5,Refes6}. These models with 
CH interaction provide us with a 
unique possibility to study unconventional mechanisms for Cooper pairing, 
metal-insulator, and insulator-superconductor transitions.

An interesting CH model with a rich ground state phase 
diagram has been proposed by Simon and Aligia \cite{SA}. The 1D version of 
the Simon-Aligia Hamiltonian reads:
%%%%%%%%%%%%%%%%%%%%%%%%%%%%%%%%%%%%%%%%%%%%%%%%%
\begin{eqnarray}\label{SAmodel1} 
{\cal H}& = & t_{{\em eh}} \sum_{n,\sigma}\hat Q_{n,n+1,\sigma}
(1-\hat \rho_{n,-\sigma})(1-\hat \rho_{n+1,-\sigma})\nonumber \\
& + &t_{{\em ed}} \sum_{n,\sigma}\hat Q_{n,n+1,\sigma}
\hat\rho_{n,-\sigma}\hat \rho_{n+1,-\sigma} \nonumber \\
&+& t_{{\em dd}}\sum_{n,\sigma}\hat Q_{n,n+1,\sigma}
(\hat \rho_{n,-\sigma} + \hat \rho_{n+1,-\sigma} 
- 2 \hat \rho_{n,-\sigma}\hat \rho_{n+1,-\sigma}) \nonumber \\ 
& + & {1 \over 2}U \sum_{n,\sigma}\hat \rho_{n,\sigma}\hat \rho_{n,-\sigma}
+ V\sum_{n}\hat \rho_{n}\hat \rho_{n+1}
\end{eqnarray}
%%%%%%%%%%%%%%%%%%%%%%%%%%%%%%%%%%%%%%%%%%%%%%%%
where $\hat \rho_{n,\sigma}= c^{\dagger}_{n, \sigma}c_{n,\sigma}$, 
$\hat \rho_{n} = \sum_{\sigma} \hat \rho_{n,\sigma}$, and 
$\hat Q_{n,n+1,\sigma} = c^{\dagger}_{n, \sigma}c_{n+1,\sigma} + 
c^{\dagger}_{n+1,\sigma}c_{n, \sigma}$. The first term interchanges 
an electron and a hole, while the second term interchanges an electron 
and a doublon (doubly occupied site) between nn sites. The 
effect of $t_{dd}$ is to destroy a doublon in the presence of a nn
hole into two electrons on nn sites, and vice versa.

It is useful to rewrite the Hamiltonian (\ref{SAmodel1}) in the standard way, 
combining the two-body and the three-body terms. As a result the Hamiltonian 
is rewritten as:
%%%%%%%%%%%%%%%%%%%%%%%%%%%%%%%%%%%%%%%%%%%%%%%%%
\begin{eqnarray}\label{SAmodel} 
{\cal H} & = & -t \sum_{n,\sigma}(c^{\dagger}_{n, \sigma}c_{n+1,\sigma} 
+ c^{\dagger}_{n+1,\sigma}c_{n, \sigma}) 
- \mu \sum_{n,\sigma}c^{\dagger}_{n, \sigma}c_{n, \sigma}\nonumber\\ 
& + & {1 \over 2}U \sum_{n,\sigma}\hat \rho_{n,\sigma}\hat \rho_{n,-\sigma} 
+ V\sum_{n}\hat \rho_{n}\hat \rho_{n+1}\nonumber\\
& + &t^{\ast}\sum_{n,\sigma}\hat Q_{n,n+1,\sigma}
(\hat \rho_{n,-\sigma} + \hat \rho_{n+1,-\sigma})\nonumber\\ 
& +  & T^{\ast}\sum_{n,\sigma}\hat Q_{n,n+1,\sigma}\hat \rho_{n,-\sigma}
\hat\rho_{n+1,-\sigma}.
\end{eqnarray}
%%%%%%%%%%%%%%%%%%%%%%%%%%%%%%%%%%%%%%%%%%%%%%%%
Here $t^{\ast}=t_{eh}-t_{dd}$ and $T^{\ast}= 2t_{dd} -t_{eh}-t_{ed}$. There 
are $ N_{e} $ particles, $ N_{0}$ sites and the band filling 
$\nu =  N_{e}/2N_{0}$ is controlled by the chemical potential $\mu$. 

The 2D version of the Hamiltonian (\ref{SAmodel1}) has been derived by Simon 
and Aligia as an effective one-band model resulting from tracing out the 
oxygen degrees of freedom in cuprates \cite{SA}. The model was studied by 
analytical and numerical methods, especially in the limit of strong 
interactions \cite{SA1,SA2,SA3,ES6,SA5,SA6,SA7,SA8,SA9,SA10}. The main 
attention was focused on the search for a superconducting ground state. Away 
from half-filling and for $t^{\ast} \simeq t$ the properties of the system are 
determined by the two-body CH term ($t^{\ast}$) and are in 
qualitative agreement with results for the standard CH model 
($T^{\ast}=0$) \cite{MH1,MH2,JM}. There is a transition into a superconducting 
phase for particular band-fillings and sufficiently small on-site repulsion 
\cite{SA1,SA10}. The effective interaction originating from the $t^{\ast}$ term
which appears in the continuum-limit theory, is given by $t^{\ast}\cos(\pi\nu)$
\cite{JM}. Therefore, in the half-filled band-case the three-body term becomes 
crucial. For $t^{\ast}=t$ and $T^{\ast}<0$ an insulator-metal transition for 
sufficiently small $U$ and $V$ has been demonstrated \cite{SA2,SA3,ES6,SA5}. 
The nature of superconducting instabilities in the metallic phase was 
investigated numerically and within a mean-field approach 
\cite{SA6,SA7,SA8,SA9}. Recently, also the possibility for realizing triplet 
superconductivity (TS) in the ground state of the model (\ref{SAmodel}) at 
half-filling was studied \cite{SA8,SA9}.

An important feature of the CH interaction is its {\it 
site-off-diagonal nature}. At half-filling this provides the principal 
possibility for realizing {\it bond located ordering} \cite{GJ1,GJ2}. In this 
paper we study the model Hamiltonian (\ref{SAmodel}) using the weak-coupling 
field-theory approach. We focus on the {\it search for bond-located ordered 
phases}. Such an ordering has not been considered in previous studies. We show 
that for $T^{\ast}>0$ the three-body interaction enhances the Peierls 
instability in the system. Near the frustration line $U=2V$ of the extended 
($U-V$) Hubbard model \cite{Emery1}, for $|U-2V|<8T^{\ast}/\pi$ and 
$V>-4T^{\ast}/\pi$ the long range ordered (LRO) dimerized ground state with 
order parameter
%%%%%%%%%%%%%%%%%
\begin{equation}\label{dimerorder1}
\Delta_{dimer} = (-1)^{n} \sum_{\sigma} \hat Q_{n,n+1,\sigma}
\end{equation}
%%%%%%%%%%%%%%%%% 
is realized. For $T^{\ast}<0$ the bond-located spin-density-wave (bd-SDW) phase
with order parameter
%%%%%%%%%%%%%%%%%%%%%%%%%%%%%%%%%%%%%%%
\begin{equation}\label{b-SDW1} 
\Delta_{\small bd-SDW} = (-1)^{n} \sum_{\sigma}\sigma \hat Q_{n,n+1,\sigma}
\end{equation}
%%%%%%%%%%%%%%%%%%%%%%%%%%%%%%%%%%%%%%% 
and the the charge-density wave ($CDW$) phase show an identical power-law 
decay of the correlation functions at large distances for 
$|U-2V|<8|T^{\ast}|/\pi$ and $V>0$. The $bd-SDW$ phase corresponds to a 
staggered magnetization located on bonds between sites.

The paper is organized as follows: In Sect. II the symmetry of the model is 
reviewed. In Sect. III the continuum-limit bosonized version of the model is 
constructed. In Sect. IV we discuss the weak-coupling phase diagram. Finally,
Sect. V is devoted to a discussion and to concluding remarks on the ground 
state phase diagram.

\section{\bf Symmetries of the model}

In the absence of the CH interaction ($t^\ast=T^\ast=0$) Eq. (\ref{SAmodel}) is
the Hamiltonian of the extended Hubbard model. The ground state phase diagram 
of the 1/2-filled extended Hubbard model is well studied 
\cite{Emery1,Voit2,EHM}: the low-energy properties of the model are essentially
determined by the parameter $U-2|V|$. The insulating ground state for $U>2|V|$ 
is dominated by spin-density wave ($SDW$) correlations. The line $U=2|V|$ 
corresponds to a Luttinger liquid (LL) phase. In the case of repulsive nn 
interaction ($V>0$), the $U=2V$ line corresponds to a transition from the SDW 
phase (for $U>2V$) into an insulating LRO $CDW$ phase for $U<2V$ 
\cite{Emery1,Voit2,EHM}. In the case of attractive nn interaction ($V<0$) the 
$U=2|V|$ line corresponds to a transition from the insulating $SDW$ phase into 
a metallic phase with dominating superconducting instabilities \cite{Voit2}. 

When the CH interaction is added to the model two new aspects 
appear. The first is the {\it site-off-diagonal character} of the 
CH coupling which provides a possibility for bond-located 
ordering. The second is the symmetry aspect. In the general case the CH 
interaction violates the electron-hole symmetry \cite{Hirsch}. This 
leads to an essential band-filling dependence of the phase diagram 
\cite{Hirsch,JM}.

Let us first consider the symmetry aspect. The three generators of the 
spin-$SU(2)$ algebra 
%%%%%%%%%%%%%%%%%%%%%%%%%%%%%%%%%%%%%%%%%%%%%%
\begin{eqnarray}\label{spin1}
S^{+} = \sum_{n}c^{\dagger}_{n,\uparrow}c_{n, \downarrow}, \ \ \ \
S^{-} = \sum_{n}c^{\dagger}_{n,\downarrow}c_{n, \uparrow},\nonumber\\
S^{z} = \sum_{n}{1 \over 2}(c^{\dagger}_{n,\uparrow}c_{n, \uparrow} - 
c^{\dagger}_{n,\downarrow}c_{n, \downarrow}) ,
\end{eqnarray}
%%%%%%%%%%%%%%%%%%%%%%%%%%%%%%%%%%%%%%%%%%%%
commute with the Hamiltonian (\ref{SAmodel}) which shows its $SU(2)$-spin 
invariance.

The electron-hole transformation 
%%%%%%%%%%%%%%%%%%%%%%%%%%%%%%%%%%%%
\begin{equation}\label{EHtransf}
c_{n, \sigma} \rightarrow (-1)^{n}c^{\dagger}_{n, \sigma},
\end{equation}
%%%%%%%%%%%%%%%%%%%%%%%%%%%%%%%%%%%%%%%%%%%%%
converts ${\cal H}\{t,U,V,t^{\ast},T^{\ast}\}\rightarrow{\cal H}\{\tilde{t},U,
V, \tilde{t}^{\ast},T^{\ast}\}$ with
%%%%%%%%%%%%%%%%%%%%%%%%%%%%%%%%%%%%%%%%%%%%
\begin{equation}\label{ehhamilt1}
\tilde{t}= t-2t^{\ast}-T^{\ast}, \ \ \ \ 
\tilde{t}^{\ast} = -t^{\ast}-T^{\ast}
\end{equation}
%%%%%%%%%%%%%%%%%%%%%%%%%%%%%%%%%%%%%%%%%%%%
and therefore the Hamiltonian (\ref{SAmodel}) does possess electron-hole 
symmetry for $2t^{\ast} +T^{\ast} =0$.

At half-filling and for $V=2t^{\ast} +T^{\ast} =0 $ the model (\ref{SAmodel}) 
is characterized by an additional important symmetry. The transformation
%%%%%%%%%%%%%%%%%%%%%%%%%%%%%%%%%%%%
\begin{eqnarray}\label{spinEHtransf}
c_{n, \uparrow} &\rightarrow& c^{\dagger}_{n, \uparrow}\nonumber\\
c_{n, \downarrow} & \rightarrow & (-1)^{n}c^{\dagger}_{n, \downarrow},
\end{eqnarray}
%%%%%%%%%%%%%%%%%%%%%%%%%%%%%%%%%%%%%%%%%%%%%
interchanges the charge and spin degrees of freedom and  converts
%%%%%%%%%%%%%%%%%%%%%%%%%%%%%%%%%%%%%%%%%%%%
\begin{equation}\label{ehhamilt2}
{\cal H}(t, U , T^{\ast}) \rightarrow {\cal H}(t, -U, T^{\ast}).
\end{equation}
%%%%%%%%%%%%%%%%%%%%%%%%%%%%%%%%%%%%%%%%%%%%
Therefore, in this case, the charge sector is governed by the same $SU(2)$ 
symmetry as the spin sector and the model has the $SU(2)\otimes SU(2)$ symmetry
\cite{ES3} with generators:
%%%%%%%%%%%%%%%%%%%%%%%%%%%%%%%%%%%%%%%%%%%%%%
\begin{eqnarray}\label{spin2}
\eta^{+} = \sum_{n}(-1)^{n}c^{\dagger}_{n,\uparrow}c^{\dagger}_{n,\downarrow}, 
\ \ \ \
\eta^{-} = \sum_{n}(-1)^{n}c_{n,\downarrow}c_{n, \uparrow},\nonumber\\
\eta^{z} = \sum_{n}{1 \over 2}(1-c^{\dagger}_{n,\uparrow}c_{n, \uparrow} - 
c^{\dagger}_{n,\downarrow}c_{n, \downarrow}).
\end{eqnarray}%
%%%%%%%%%%%%%%%%%%%%%%%%%%%%%%%%%%%%%%%%%%%

For the half-filled Hubbard model the $SU(2) \otimes SU(2)$ symmetry implies 
that the gapful charge and the gapless spin sectors for $U>0$ are mapped by the
transformation Eq. (\ref{spinEHtransf}) into a gapful spin and a gapless charge
sector for $U<0$. Moreover, at $U<0$ the model is characterized by the 
coexistence of $CDW$ and singlet superconducting $(SS)$ instabilities in the 
ground state \cite{FK}.

Contrary to the on-site Hubbard interaction $U$ the $T^{\ast}$ term remains 
invariant with respect to the transformation Eq. (\ref{spinEHtransf}). This 
immediately implies that for a given 
$T^{\ast}$ and 
\begin{itemize}
\item
for $U=0$ the properties of the charge and the spin sectors are identical;
\item
for $U \neq 0$ there exists a critical value of the Hubbard coupling $U_{c}$ 
corresponding to a crossover from the $T^{\ast}$ dominated regime into a $U$ 
dominated regime.
\item
The LL parameters of the model characterizing the gapless charge ($K_{c}$) and 
spin ($K_{s}$) degrees of freedom are $K_{c}=K_{s}=1$.
\end{itemize}

For nonzero nn interaction $(V \neq 0)$ the spin $SU(2)$--symmetry remains 
unchanged, while the symmetry of the charge sector is reduced to a 
$U(1)$--symmetry (conservation of charge). In this case the gapless charge 
sector is parametrized by a fixed-point value of the parameter 
$K_{c}=K_{c}^{\ast}$ which essentially depends on the bare values of the
coupling constants. This results in a different power-law decay at large 
distances for density-density and superconducting correlations, supporting 
$CDW$ for $V> 0$ and superconductivity for $V< 0$. However, due to the 
$SU(2)$--spin symmetry the dynamical generation of a gap in the spin 
excitation spectrum supports $SS$ superconductivity. In the case of a 
gapless spin sector both $SS$ and $TS$ correlations show an identical 
power-law decay at large distances.

\section{\bf Continuum-limit theory and bosonization.}

In this section we construct the continuum-limit version of the model Eq.
(\ref{SAmodel}) at half-filling. While this procedure has a long history and 
is reviewed in many places \cite{GNT}, for clarity we briefly sketch the 
most important points.

The field theory treatment of 1D systems of correlated electrons is based on 
the weak-coupling approach $|U|, |V|, |t^{\ast}|, |T^{\ast}| \ll t$. Assuming 
that the low energy physics is controlled by states near the Fermi points 
$ \pm k_{F}$ ($k_{F} =\pi/2a_{0}$, where $a_{0}$ is the lattice spacing) we 
linearize the spectrum around these points and obtain two species (for each 
spin projection $\sigma$) of fermions, $R_{\sigma}(n)$ and $L_{\sigma}(n)$, 
which describe excitations with dispersion relations $E = \pm v_{F}p$. Here, 
$v_{F}=2 ta_{0}$ is the Fermi velocity and the momentum $p$ is measured from 
the two Fermi points. More explicitly, one decomposes the momentum expansion 
for the initial lattice operators into two parts centered around $\pm k_{F}$ to
obtain the mapping:
%%%%%%%%%%%%%%%%%%%%%%%%%%%%%%%%%%%%%%%%%%%%%
\begin{equation}\label{linearization}
c_{n,\sigma} \rightarrow  {\it i}^{n}R_{\sigma}(n) + 
(-{\it i})^{n}L_{\sigma}(n),
\end{equation}
%%%%%%%%%%%%%%%%%%%%%%%%%%%%%%%%%%%%%%%%%%%%%%%%
where the fields $R_{\sigma}(n)$ and $L_{\sigma}(n)$ describe 
right-moving and left-moving particles, respectively, and are assumed to be 
smooth on the scale of the lattice spacing. This allows us to introduce the 
continuum fields $R_\sigma(x)$ and $L_\sigma(x)$ by
%%%%%%%%%%%%%%%%%%%%%%%%%%%%%%%%%%%%%%%%%%%%%
\begin{eqnarray}\label{continuumfields}
R_{\sigma}(n) & \rightarrow & \sqrt{a_{0}}R_{\sigma}(x=na_{0}),\nonumber\\
L_{\sigma}(n) & \rightarrow & \sqrt{a_{0}}L_{\sigma}(x=na_{0}).
\end{eqnarray}
%%%%%%%%%%%%%%%%%%%%%%%%%%%%%%%%%%%%%%%%%%%%%%

In terms of the continuum fields the free Hamiltonian reads:
%%%%%%%%%%%%%%%%%%%%%%%%%%%%%%%%%%%%%%%%%%%%%
\begin{equation}\label{freelinearized}
{\cal H}_{0}  =  E_{0} - iv_{F}\sum_\sigma \int  dx 
[:R^{\dagger}_{\sigma}\partial_{x} R_{\sigma}: - 
:L^{\dagger}_{\sigma}\partial_{x}L_{\sigma}:]
\end{equation}
%%%%%%%%%%%%%%%%%%%%%%%%%%%%%%%%%%%%%%%%%%%%
which is recognized as the Hamiltonian of a free massless Dirac field and the 
symbols :...: denote normal ordering with respect to the ground state of the 
free system.

The advantage of the linearization of the spectrum is twofold: the initial 
lattice problem is reformulated in terms of smooth continuum fields and -- 
using the bosonization procedure -- is mapped to the theory of two decoupled 
quantum sine-Gordon (SG) models describing charge and spin degrees of freedom, 
respectively. 

In terms of the continuum fields the initial lattice operators have the form
%%%%%%%%%%%%%%%%%%%%%%%%%%%%%%%%%%%%%%%%%%%%%
\begin{eqnarray}
\hat\rho_{n,\sigma}-\frac{1}{2}&\equiv& :\hat\rho_{n,\sigma}:=\nonumber\\
a_{0}\{(:R^{\dagger}_{\sigma}(x)R_{\sigma}(x): &+& 
:L^{\dagger}_{\sigma}(x)L_{\sigma}(x):)\nonumber\\
+  (-1)^{n}(R^{\dagger}_{\sigma}(x)L_{\sigma}(x) &+&
L^{\dagger}_{\sigma}(x)R_{\sigma}(x))\},\label{mapping2}\\
:\hat Q_{n,n+1;\sigma}:\equiv\hat Q_{n,n+1;\sigma}&-&\frac{2}{\pi}=\nonumber\\
2a_{0}{\it i}(-1)^{n}(R^{\dagger}_{\sigma}(x)L_{\sigma}(x)&-& 
L^{\dagger}_{\sigma}(x)R_{\sigma}(x)).\label{mapping3}
\end{eqnarray}
%%%%%%%%%%%%%%%%%%%%%%%%%%%%%%%%%%%%%%%%%

The second step is to use the standard bosonization expressions for fermionic 
bilinears \cite{LutEm}:
%%%%%%%%%%%%%%%%%%%%%%%%%%%%%%%%%%%%%%%%
\begin{eqnarray}
-i [:R^{\dagger}_{\sigma}\partial_{x} R_{\sigma}:&-& 
:L^{\dagger}_{\sigma}\partial_{x}L_{\sigma}:] \rightarrow\nonumber\\
{1\over 2}\{P^2_{\sigma}(x)&+&(\partial_{x}\varphi_{\sigma})^2\},\label{bos1}\\
:R^{\dagger}_{\sigma}(x)R_{\sigma}(x):&+& 
:L^{\dagger}_{\sigma}(x)L_{\sigma}(x): \rightarrow 
{1 \over {\sqrt{\pi}}}\partial_{x}\phi_{\sigma}(x),\label{bos2}\\
:R^{\dagger}_{\sigma}(x)R_{\sigma}(x):&-& 
:L^{\dagger}_{\sigma}(x)L_{\sigma}(x): \rightarrow  
- {1 \over {\sqrt{\pi}}} P_{\sigma}(x),\label{bos3}\\
R^{\dagger}_{\sigma}(x)L_{\sigma}(x)&\rightarrow &  
-{i \over \sqrt{2 \pi a_{0}}}\exp(- i\sqrt{4\pi} 
\phi_{\sigma}(x)).\label{bos4}
\end{eqnarray}
%%%%%%%%%%%%%%%%%%%%%%%%%%%%%%%%%%%%%%%%
We thereby obtain
%%%%%%%%%%%%%%%%%%%%%%%%%%%%%%%%%%%%%%%%%%%%%
\begin{eqnarray}
:\hat\rho_{n, \sigma}: \rightarrow
a_{0}\{\frac{1}{\sqrt{\pi}}\partial_{x} 
\varphi_{\sigma}-(-1)^{n} \frac{1}{\pi a_{0}}
\sin(\sqrt{4\pi}\varphi_{\sigma})\}&,&\label{mapping2.a}\\
:\hat Q_{n,n+1;\sigma} : \rightarrow (-1)^{n} \frac{2}{\pi} 
\cos(\sqrt{4\pi}\varphi_{\sigma})&,&\label{mapping3.a}\\
:\hat\rho_{n, \sigma}::\hat\rho_{n+1, \sigma}: \rightarrow 
\frac{2}{\pi^{2}}\sin(\sqrt{4\pi}\varphi_{\sigma})
+a^{2}_{0}\frac{2}{\pi}(\partial_{x} \varphi_{\sigma})^{2}&.&\label{mapping4}
\end{eqnarray}
%%%%%%%%%%%%%%%%%%%%%%%%%%%%%%%%%%%%%%%%%
Here, $\varphi_{\sigma = \uparrow, \downarrow}(x)$ and 
$P_{\sigma =\uparrow, \downarrow}(x)$ are a scalar field and its conjugate 
momentum, respectively, related to the spin up and spin down subsystems. In 
deriving Eq. (\ref{mapping4}) the following operator product expansion 
relations have been used:
%%%%%%%%%%%%%%%%%%%%%%%%%%%%%%%%%%%%%%%%
\begin{eqnarray}
:\partial_{x}\varphi(x)&:&:\sin(\sqrt{4\pi}\varphi(x+a_{0})): =\nonumber\\ 
\frac{1}{\sqrt{\pi}a_{0}}&:&\cos(\sqrt{4\pi}\varphi(x)):,\label{OPE1}\\
:\sin(\sqrt{4\pi}\varphi(x))&:&:\sin(\sqrt{4\pi}\varphi(x+a_{0})):=\nonumber\\
- a^{2}_{0}\pi:(\partial_{x} \varphi(x))^{2}&:& - 
\frac{1}{2}:\cos(\sqrt{16\pi}\varphi(x)):.\label{OPE2}
\end{eqnarray}
%%%%%%%%%%%%%%%%%%%%%%%%%%%%%%%%%%%%%%%%%

Finally, introducing the bosonic charge ($\varphi_{c}$) and spin 
($\varphi_{s}$) fields
%%%%%%%%%%%%%%%%%%%%%%%%%%%%%%%%%%%%%%%%
\begin{eqnarray}
\varphi_{c}&=&\frac{1}{\sqrt{2K_{c}}}(\varphi_{\uparrow}+\varphi_{\downarrow}),
\hskip0.3cm P_{c}=\sqrt{{K_{c}\over 2}}(P_{\uparrow}+P_{\downarrow}),
\label{bosecharge1}\\
\varphi_{s}&=&\frac{1}{\sqrt{2K_{s}}}(\varphi_{\uparrow}-\varphi_{\downarrow}),
\hskip0.3cm P_{s}=\sqrt{{K_{s}\over 2}}(P_{\uparrow}-P_{\downarrow}),
\label{bosespin}
\end{eqnarray}
%%%%%%%%%%%%%%%%%%%%%%%%%%%%%%%%%%%%%%%%%
and converting $\sum_{n}a_{0}\rightarrow\int dx $ we rewrite the model 
Hamiltonian in terms of two decoupled quantum SG theories, ${\cal H}=
{\cal H}_{c}+{\cal H}_{s}$, where
\begin{eqnarray}
{\cal H}_{c(s)}&=&v_{c(s)}\int dx\Big\{{1\over 2}[P^2_{c(s)}(x)+(\partial_x
\varphi_{c(s)})^2]\nonumber\\
&+& \frac{m_{c(s)}}{a_0^2}\cos(\sqrt{8\pi K_{c(s)}}\varphi_{c(s)})
\Big\},
\label{SG}
\end{eqnarray}
Here we have defined
%%%%%%%%%%%%%%%%%%%%%%%%%%%%%%%%%%%%%%%%%%%%%%
\begin{eqnarray}
K_{c}&=&(1+g_{c})^{1/2}\simeq 1+{1\over 2}g_{c},\hskip0.3cm
m_{c}=\frac{g_{u}}{2\pi},\label{Kc}\\
K_{s}&=&(1+g_{s})^{1/2}\simeq 1+{1\over 2}g_{s},\hskip0.3cm
m_{s}=\frac{g_{\perp}}{2 \pi },\label{Ks}\\
v_{c}&=&\frac{v_{F}}{K_{c}} \simeq (1 - {1 \over 2}g_{c}),\hskip0.3cm 
v_{s}= \frac{v_{F}}{K_{s}} \simeq v_{F}(1 - {1 \over 2}g_{s}).\label{McMs}
\end{eqnarray}
%%%%%%%%%%%%%%%%%%%%%%%%%%%%%%%%%%%%%%%%%%%%%%%%
The small dimensionless coupling constants are given by:
%%%%%%%%%%%%%%%%%%%%%%%%%%%%%%%%%%%%%%%%%%%%%%%
\begin{eqnarray}
g_{c} &= & - {1 \over 2 \pi t}(U+6V+ 8T^{\ast}/ \pi ),\label{gu}\\
g_{u} &= & - {1 \over 2\pi t}(U - 2V + 8T^{\ast}/ \pi),\label{gc}\\
g_{s} = g_{\perp}  & =  &{1 \over 2\pi t}(U- 2V - 8T^{\ast}/ \pi)\label{gs}.
\end{eqnarray}
%%%%%%%%%%%%%%%%%%%%%%%%%%%%%%%%%%%%%%%%%%%%%%%
The relation between $K_{c}$ ($K_{s}$), $m_{c}$ ($m_{s}$), and $g_{c}$ 
($g_{s}$), $g_{u}$ ($g_{\perp}$) is universal in the weak coupling limit. 

In obtaining (\ref{SG}) only nonoscillating terms of the order 
$\sim a_{0}$ have been kept. In addition, strongly irrelevant terms 
$\sim$ $\cos(\sqrt{8\pi K_{c}}\varphi_{c})\cos(\sqrt{8\pi K_{s}}\varphi_{s})$ 
describing umklapp scattering processes with parallel spins were omitted. 

The mapping of the initial lattice Hamiltonian Eq. (\ref{SAmodel}) into the 
continuum theory of quantum SG models Eq. (\ref{SG}) allows to study the ground
state phase diagram of the system based on the infrared properties of the SG 
Hamiltonians. The corresponding behavior of the SG model is described by the 
pair of renormalization group equations for the effective coupling constants 
$\Gamma_{i}$ \cite{Wieg}  
%%%%%%%%%%%%%%%%%%%%%%%%%%%%%%%%%%
\begin{eqnarray}
d\Gamma_{u}/dL &=& - \Gamma_{c} \Gamma_{u} , \nonumber \\ 
d\Gamma_{c}/dL &=& - \Gamma^{2}_{u},\label{RGsg.a}\\ 
d\Gamma_{\perp}/dL & = &- \Gamma_{s} \Gamma_{\perp}, \nonumber \\ 
d\Gamma_{s}/dL& = & - \Gamma^{2}_{\perp},\label{RGsg.b}  
\end{eqnarray}
%%%%%%%%%%%%%%%%%%%%%%%%%%%%%%%%%%%%%
where $L=\ln{(a_0)} $ and $\Gamma_{i}(0) = g_{i} $. Each pair of equations 
(\ref{RGsg.a}) and (\ref{RGsg.b}) describes the Kosterlitz--Thouless 
transition \cite{KT} in the charge and spin channels. The flow lines lie on 
the hyperbolae  
%%%%%%%%%%%%%% 
\begin{equation}\label{flowlines} 
\Gamma_{c (s)}^{2} - \Gamma_{u (\perp)}^{2} = \mu^{2}_{c (s)} = 
g_{c (s)}^{2} - g_{u (\perp)}^{2}, 
\end{equation} 
%%%%%%%%%%%%%%%%%%%% 
and -- depending on the relation between the bare coupling constants 
$g_{c(s)}$ and $g_{u(\perp)}$ -- exhibit two different regimes:

For $g_{c}\geq |g_{u}|$ ($g_{s}\geq |g_{\perp}|$) we are in the weak coupling 
regime; the effective mass $M_{c (s)} \rightarrow 0$. The 
low energy (large distance) behavior of the gapless charge (spin) degrees of 
freedom is described by a free scalar field
%%%%%%%%%%%%%%%%%%%%%%%%%%%%%%%%
\begin{equation}\label{KG}
{\cal H}_{c (s)} ={1 \over 2} v_{c (s)} \int dx 
\{(\partial_{x} \theta_{c (s)})^{2} + 
(\partial_{x} \varphi_{c (s)})^{2}\} 
\end{equation}
%%%%%%%%%%%%%%%%%%%%%%%%%%%%%%%%%%%%%%%% 
where $\partial_{x}\theta_{c (s)}=P_{c (s)}$.

The corresponding correlations show a power law decay
%%%%%%%%%%%%%%%%%%%%%%%%%%%%%%%%%%%%%%%%%%%%%%%
\begin{eqnarray}
\langle e^{i\sqrt{2\pi K}\varphi(x)} e^{-i \sqrt{2\pi K} \varphi (x')} 
\rangle \sim \left| x - x' \right|^{- K},\label{freecorrelations1}\\
\langle e^{i \sqrt{2\pi/K}\theta(x)} e^{-i\sqrt{2\pi/K}
\theta(x')}\rangle 
 \sim \left| x - x' \right| ^{-1 / K},\label{freecorrelations2}
\end{eqnarray}
%%%%%%%%%%%%%%%%%%%%%%%%%%%%%%%%%%%%%%%%%%%%%%%%
and the only parameter controlling the infrared behavior in the gapless regime 
is the fixed-point value of the effective coupling constants $K_{c(s)}$. 

For $g_{c}<|g_{u}|$ ($g_{s}<|g_{\perp}|$) the system scales to the strong 
coupling regime; depending on the sign of the bare mass $m_{c(s)}$ the 
effective mass $M_{c(s)}\rightarrow\pm\infty$, which signals the crossover to 
the strong coupling regime and indicates the dynamical generation of a 
commensurability gap in the charge (spin) excitation spectrum. The fields 
$\varphi_{c}$ ($\varphi_{s}$) get ordered with the vacuum expectation values 
\cite{ME}  
%%%%%%%%%%%%%%%%%% 
\begin{equation}\label{orderfields}
\langle\varphi_{c(s)}\rangle =\left\{ \begin{array}{l@{\quad}}\sqrt{
\displaystyle{\pi\over 8K_{c(s)}}} \hskip0.5cm(m_{c(s)}>0) \\ 0 \hskip1.5cm
(m_{c(s)}<0)\end{array}\right. \, . 
\end{equation} 
%%%%%%%%%%%%%%%%%%%

\section{Phase diagram}

Let us now consider the weak-coupling ground state phase diagram of the model 
Eq. (\ref{SAmodel}). To clarify the symmetry properties of the various ground 
states of the system we use the usual order parameters describing the short 
wavelength fluctuations of the {\it site}-located charge-density,
%%%%%%%%%%%%%%%%%%%%%%%%%%%%%%%%%%%%%%%%
\begin{eqnarray}\label{CDWop}
\Delta_{\small CDW} & = & (-1)^{n} \sum_{\sigma}\rho_{n, \sigma}\nonumber\\
& \sim & \sin(\sqrt{2\pi K_{c}}\varphi_{c}) 
\cos( \sqrt{2\pi K_{s}}\varphi_{s}) \, ,
\end{eqnarray}  
%%%%%%%%%%%%%%%%%%%%%%%%%%%%%%%%%%%% 
the {\it site}-located spin-density  
%%%%%%%%%%%%%%%%%%%%%%%%%%%%%%%%%%%%%
\begin{eqnarray}\label{SDWop}   
\Delta_{\small SDW} & = &\sum_{\sigma}\sigma \rho_{n, \sigma}\nonumber\\
& \sim & \cos(\sqrt{2\pi K_{c}}\varphi_{c})
\sin(\sqrt{2\pi K_{s}}\varphi_{s}) \, ,
\end{eqnarray}
%%%%%%%%%%%%%%%%%%%%%%%%%%%%%%%%%%%%%
%%%%%%%%%%%%%%%%%%%%%%%%%%%%%%%%%%%%%%%
and two superconducting order parameters corresponding to singlet 
($\Delta_{SS}$) and triplet ($\Delta_{TS}$) superconductivity:
%%%%%%%%%%%%%%%%%%%%%%%%%%%%%%%%%%%%%%%%
\begin{eqnarray}
\Delta_{SS}(x) & = &R^{\dagger}_{\uparrow}(x)L^{\dagger}_{\downarrow}(x)  
- R^{\dagger}_{\downarrow}(x)L^{\dagger}_{\uparrow}(x) \nonumber\\
& \sim & \exp(i \sqrt{\frac{2\pi}{K_{c}}}\theta_{c}) 
\cos(\sqrt{2 \pi K_{s}}\varphi_{s}),\label{SSop}\\
\Delta_{TS}(x) & = & R^{\dagger}_{\uparrow}(x)L^{\dagger}_{\downarrow}(x)  + 
R^{\dagger}_{\downarrow}(x)L^{\dagger}_{\uparrow}(x) \nonumber\\
& \sim & \exp(i  \sqrt{\frac{2\pi}{K_{c}}}\theta_{c}) 
\sin(\sqrt{2 \pi K_{s}}\varphi_{s}).\label{TSop}
\end{eqnarray}
%%%%%%%%%%%%%%%%%%%%%%%%%%%%%%%%%%%%%%%%%%%%%%
In adition we use a set of order parameters \cite{NERS} describing the short 
wavelength fluctuations of the {\it bond}-located charge-- and spin--density  
%%%%%%%%%%%%%%%%%%%% 
\begin{eqnarray}
\Delta_{\small dimer} &  = &  (-1)^{n} \sum_{\sigma}
\hat Q_{n,n+1,\sigma}\nonumber\\ 
& \sim  & \cos(\sqrt{2\pi K_{c}}\varphi_{c}) 
\cos(\sqrt{2\pi K_{s}}\varphi_{s}),\label{b-CDW}\\ 
\Delta_{\small bd-SDW} &  = &  (-1)^{n} \sum_{\sigma}\sigma 
\hat Q_{n,n+1,\sigma}\nonumber\\ 
& \sim  & \sin(\sqrt{2\pi K_{c}}\varphi_{c})
\sin(\sqrt{2\pi K_{s}}\varphi_{s}).\label{b-SDW} 
\end{eqnarray}
%%%%%%%%%%%%%%%%%%%%%%%%%%%%%%%%%%%%%%%

With the results of the previous section for the excitation spectrum and the 
behavior of the corresponding fields Eqs. 
(\ref{freecorrelations1})--(\ref{orderfields}) we now analyze the ground state 
phase diagram.

\subsection{The $SU(2) \otimes SU(2)$ symmetric case}

We first consider the $SU(2)\otimes SU(2)$ symmetric case for 
$2t^{\ast}+T^{\ast}= V =0$. In this case the coupling constants parametrizing 
the charge and spin degrees of freedom are given by
%%%%%%%%%%%%%%%%%%%%%%%%%%%%%%%%%%%%%%%%%%%%%%%
\begin{eqnarray}
g_{c} = g_{u} &= & - {1 \over \pi v_{F}}(U + 8T^{\ast} / \pi),\nonumber\\
g_{s} =g_{\perp} &= &  {1 \over \pi v_{F}}(U - 8T^{\ast}/ \pi).\label{param}
\end{eqnarray}
%%%%%%%%%%%%%%%%%%%%%%%%%%%%%%%%%%%%%%%%%%%%%%%
Although the given parameters are determined within the weak-coupling 
approach ($|g_{i}| \ll 1$) the relations Eqs. (\ref{param}) are universal and 
determined {\it by the symmetries of the model only}. This strongly restricts 
the scaling trajectories along the separatrix $\mu = 0 $ (see Fig. 1). 
The $SU(2) \otimes SU(2)$ is easily seen from Eqs. (\ref{param}): each 
channel is characterized by one parameter $g_{c}$ and $g_{s}$, respectively, 
and the electron-hole transformation Eq. (\ref{spinEHtransf}) only 
interchanges the bare values of these parameters.

\begin{figure}
\mbox{\epsfxsize 8.0cm\epsffile{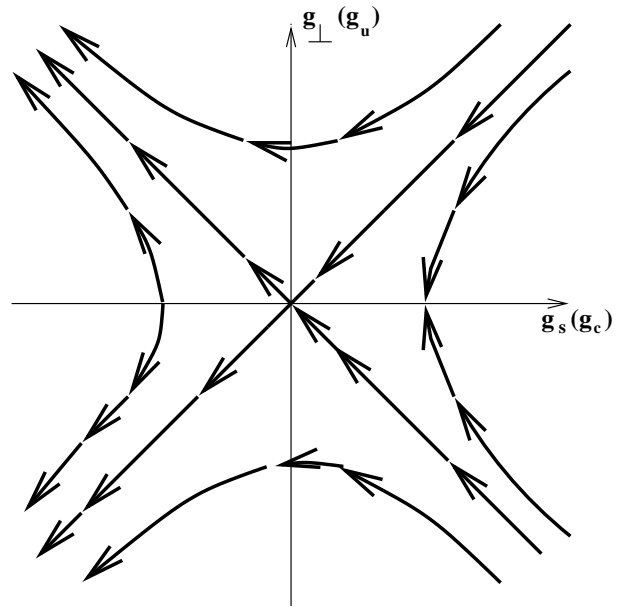}}
\vskip0.5cm
\caption[]{
The renormalization--group flow diagram; the arrows denote the 
direction of flow with increasing length scale.}
\label{fig1}
\end{figure} 

It follows from (\ref{param}) that for $U > -8T^{\ast}/ \pi$ there is a gap 
in the charge excitation spectrum, and the $\varphi_{c}$ field is ordered 
with vacuum expectation value $\langle \varphi_{c} \rangle = 0$. In the 
weak-coupling regime for $U \leq - 8T^{\ast}/ \pi$ where 
$M_{c} \rightarrow 0$, gapless charge excitations are described by the free 
bose field with the fixed-point value $K^{\ast}_{c}=1$.

The spin sector is massive for $U < 8T^{\ast}/ \pi$. The dynamical generation 
of a gap in the spin sector is accompanied by the ordering of the 
$\varphi_{s}$ field with vacuum expectation value 
$\langle\varphi_{s}\rangle=0$. For 
$U > 8T^{\ast}/ \pi$ the spin sector is gapless, $M_{s} \rightarrow 0$ and 
the fixed-point value $K^{\ast}_{s}=1$.

\begin{figure}
\mbox{\epsfxsize 8.3cm\epsffile{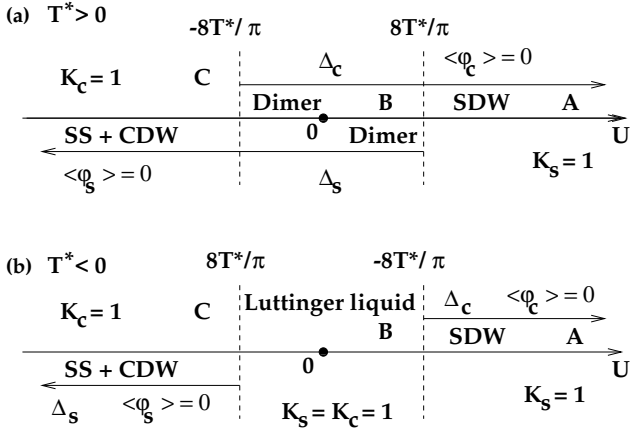}}
\vskip0.5cm
\caption[]{
Phase diagram of the model Hamiltonian Eq. (2) for the case of a 
half-filled band: (a) $V=0$, $T^{\ast}>0$ and (b) $T^{\ast}<0$.}
\label{fig2}
\end{figure} 

There are three different sectors in the phase diagram. We start with a 
discussion of the $T^{\ast}>0$ case (see Fig. 2a). 

{\bf Sector A}: $U \geq 8T^{\ast}/ \pi$.
There is a gap in the charge excitation spectrum. The charge field is ordered 
$\langle \varphi_{c} \rangle =0$. The spin excitation spectrum is gapless. 
The fixed-point value of the parameter $K_s^\ast=1$. Using Eqs. 
(\ref{CDWop})-(\ref{b-SDW}) and 
(\ref{freecorrelations1})-(\ref{freecorrelations2}) one obtains that the 
superconducting and $CDW$ instabilities are suppressed. The $SDW$ and $Dimer$ 
correlations show an identical power-law decay at large distances 
%%%%%%%%%%%%%%%%%%%%%%%%%%%%%%%%%%%%%%%%%%%
\begin{eqnarray}\label{correl3}
\langle\Delta_{\small SDW}(x)\Delta_{\small SDW}(x')\rangle &= & 
\langle \Delta_{\small dimer}(x) \Delta_{\small dimer}(x')\rangle \nonumber\\
& \sim & \left| x - x' \right|^{-1}.
\end{eqnarray}
%%%%%%%%%%%%%%%%%%%%%%%%%%%%%%%%%%%%%%%%%%
The coexistence of the $SDW$ and $Dimerization$ instabilities in the 
repulsive Hubbard model is the mechanism for the Spin-Peierls transition at 
$U \gg t$.

{\bf Sector B}: $8T^{\ast}/ \pi >U >- 8T^{\ast}/ \pi$. 
For $U<8T^{\ast}/\pi$ a spin gap opens. The charge and spin channels are 
gapped and both, charge and spin fields are ordered, 
$\langle\varphi_{c}\rangle = \langle \varphi_{s} \rangle =0$. In this case 
the LRO dimerized phase 
%%%%%%%%%%%%%%%%%%%%%%%%%%%%%%%%%%%%%%%%%%
\begin{equation}\label{correl2}
\langle \Delta_{\small dimer}(x) \Delta_{\small dimer}(x')\rangle \sim 
{\it constant}
\end{equation}
%%%%%%%%%%%%%%%%%%%%%%%%%%%%%%%%%%%%%%%%%%
is realized in the ground state.

{\bf Sector C}:  $U \leq -8T^{\ast}/ \pi$.
At $U = -8T^{\ast}/ \pi$ the charge gap closes. For $U \leq -8T^{\ast}/ \pi$ 
the phase diagram is similar to that of the half-filled attractive Hubbard 
model, i.e. there is a gap in the spin excitation spectrum. Due to the 
$SU(2)$-spin symmetry the vacuum expectation value 
$\langle \varphi_{s} \rangle = 0$. The $SDW$ and $TS$ fluctuations are 
completely  suppressed. The charge excitation spectrum is gapless and the 
fixed-point value of the parameter $K_{c}$ (due to the $SU(2)$-charge 
symmetry) is $K^{\ast}_{c}=1$. The $CDW$, $SS$, and $Dimer$ correlations show 
an identical power-law decay at large distances
%%%%%%%%%%%%%%%%%%%%%%%%%%%%%%%%%%%%%%%%%%%
\begin{eqnarray}
\langle \Delta_{\small CDW}(x)\Delta_{\small CDW}(x')\rangle &=& 
\langle \Delta_{\small SS}(x)\Delta_{\small SS}(x')\rangle =\nonumber\\
\langle \Delta_{\small dimer}(x) \Delta_{\small dimer}(x')\rangle &\sim& 
\left| x - x' \right|^{-1}.
\label{correl1}
\end{eqnarray}
%%%%%%%%%%%%%%%%%%%%%%%%%%%%%%%%%%%%%%%%%%

We next consider the case $T^{\ast}<0$ (see Fig. 2b). The spin gap regime is 
realized for $U<-8\left|T^{\ast}\right|/\pi<0$ and the charge gap regime for 
$U>8\left|T^{\ast}\right|/\pi>0$. Therefore, the properties of the model in 
the ${\bf sectors A_{1}}$ ($U>8\left|T^{\ast}\right|/\pi$) and ${\bf C_{1}}$ 
($U<-8\left|T^{\ast}\right|/\pi$) are the same as in the corresponding 
{\bf A} and {\bf C} sectors in the case of $T^{\ast}>0$.

However, for $-8\left|T^{\ast}\right|/\pi<U<8\left|T^{\ast}\right|/\pi$ both, 
the charge and the spin channel are gapless. The fixed point values of the LL 
parameters are given by
$$
K^{\ast}_{c}=K^{\ast}_{s}=1.
$$
Using Eqs. (\ref{CDWop})-(\ref{b-SDW}) and 
(\ref{freecorrelations1})-(\ref{freecorrelations2}) one obtains that 
{\it all correlations show an identical $\sim \left| x - x' \right|^{-2}$ 
decay at large distances} in this case.

\subsection{Effects of $V$}

Let us now consider the weak coupling phase diagram of the model Eq.  
(\ref{SAmodel}) for $V\neq 0$. From Eqs. (\ref{gu})-(\ref{gs}) one obtains 
that there is a gap in the spin excitation spectrum for $U<2V+8T^\ast/\pi$. The
charge excitation spectrum is gapped for $U>2|V|-8 T^\ast/\pi$ and -- in the 
case of repulsive nn interaction 
($V>0$) -- for $U < 2V - 8 T^{\ast}/ \pi$. The line $U=2|V|-8T^\ast/\pi$ 
corresponds to a metallic LL phase. This determines five different sectors in the phase diagram.

We start with the $T^{\ast}  > 0$ case (see Fig.3 a).

{\bf Sector A}: $U > 2V + 8 T^{\ast}/ \pi$.
This is the sector dominated by on-site repulsion. The roperties of the system 
in this sector coincide with that of sector A for the $V=0$ case.
The line $U=2V+8 T^{\ast}/\pi$ marks the transition into the spin 
gap regime.

{\bf Sector B}: $|U-2V| < 8 T^{\ast}/ \pi $, $V > -4 T^{\ast}/ \pi$. In this 
sector both channels are massive and the LRO dimerized phase is realized.
The line$U = 2V-8T^{\ast}/ \pi $ corresponds to a nonmagnetic metallic 
phase. Along this line the charge gap is zero, while the spin gap remains
finite, and $K_{c} \simeq 1-2V/\pi t<1$. The $TS$ instability is 
suppressed, while the $SS$ correlations
%%%%%%%%%%%%%%%%%%%%%%%%%%%%%%%%%%%%%%%%%%%
\begin{equation}\label{SS}
\langle \Delta_{\small SS}(x)\Delta_{\small SS}(x')\rangle  
\sim  \left| x - x' \right|^{-1/K_{c}}
\end{equation}
%%%%%%%%%%%%%%%%%%%%%%%%%%%%%%%%%%%%%%%%%% 
decay at large distances faster than the density-density correlations
%%%%%%%%%%%%%%%%%%%%%%%%%%%%%%%%%%%%%%%%%%
\begin{eqnarray}
\langle \Delta_{\small CDW}(x)\Delta_{\small CDW}(x')\rangle  &=& 
\langle \Delta_{\small dimer}(x) \Delta_{\small dimer}(x')\rangle \nonumber\\ 
& \sim & \left| x - x' \right| ^{-K_{c}}.
\end{eqnarray}
%%%%%%%%%%%%%%%%%%%%%%%%%%%%%%%%%%%%%%%%%

{\bf Sector ${\bf C}_{{\bf 1}}$}: $U < 2V- 8 T^{\ast}/ \pi $ and $V>0$.
A gap in the charge excitation spectrum opens once again; however, in this 
sector the vacuum expectation value of the charge field is 
$\langle \varphi_{c} \rangle = \sqrt{\pi /8K_{c}}$. The spin sector is gapped 
with $\langle \varphi_{s} \rangle = 0$. Using Eqs. (\ref{CDWop})-(\ref{b-SDW}) 
one obtains a LRO $CDW$ in the ground state:
%%%%%%%%%%%%%%%%%%%%%%%%%%%%%%%%%%%%%%%%%%
\begin{equation}\label{sectorc1}
\langle \Delta_{\small CDW}(x) \Delta_{\small CDW}(x')\rangle 
\sim  {\it constant}.
\end{equation}
%%%%%%%%%%%%%%%%%%%%%%%%%%%%%%%%%%%%%%%%%%

\begin{figure}
\mbox{\epsfxsize 8.3cm\epsffile{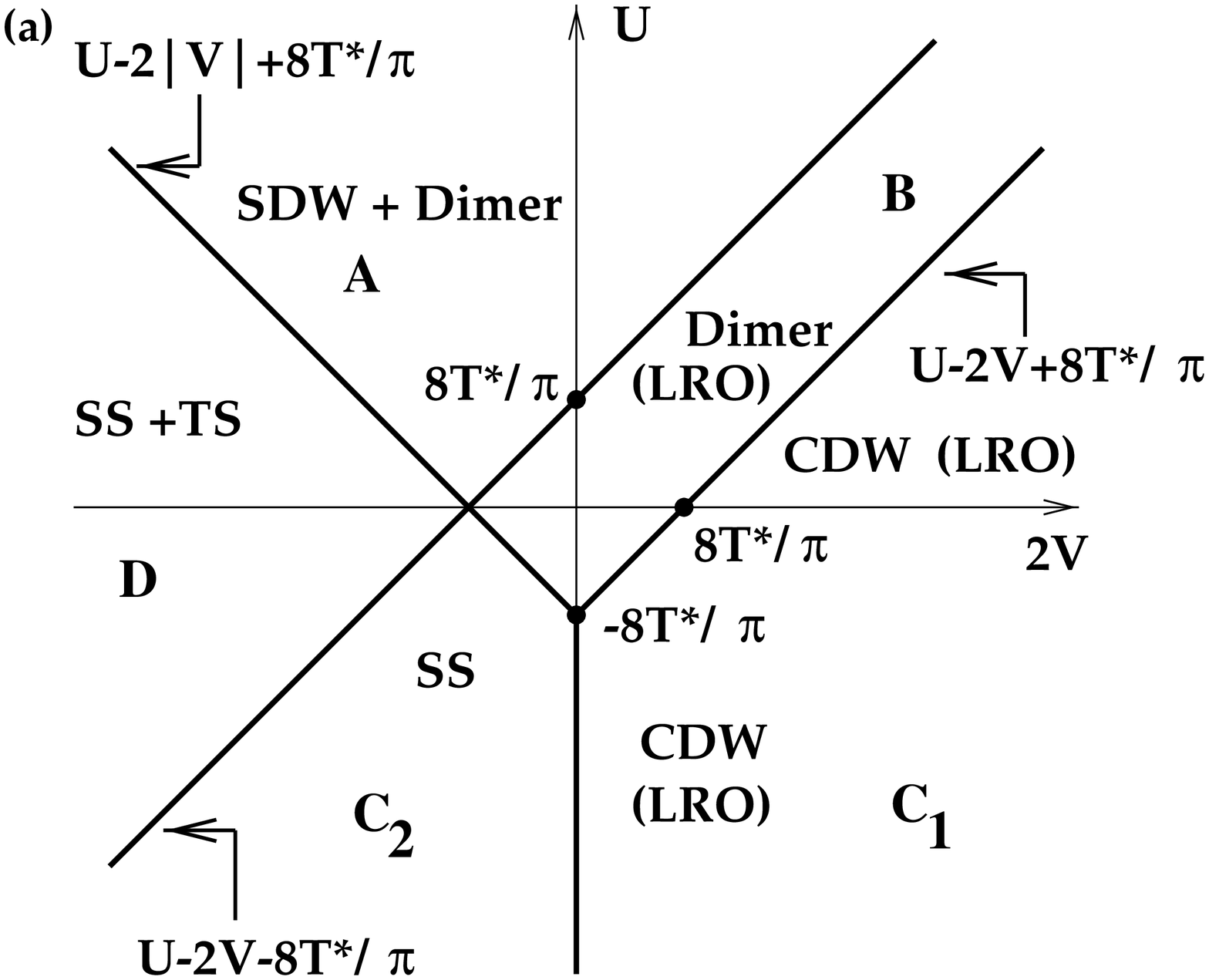}}
\vskip0.8cm
\mbox{\epsfxsize 8.3cm\epsffile{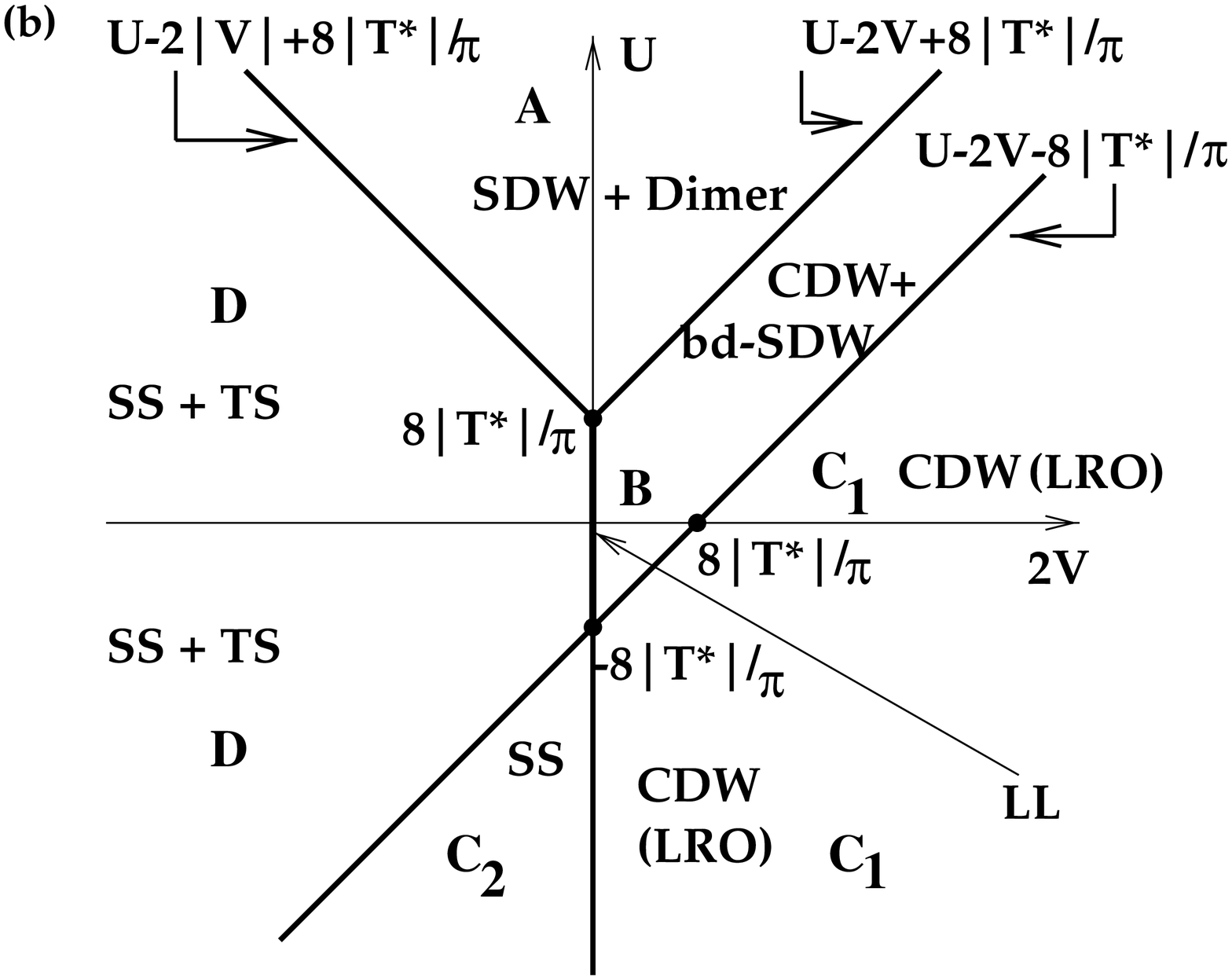}}
\vskip0.5cm
\caption[]{
Phase diagram of the model Hamiltonian Eq. (2) for the case of a 
half-filled band and (a) $T^{\ast}>0$, (b) $T^{\ast}<0$.}
\label{fig3}
\end{figure} 

In the case of a repulsive nn interaction ($V>0$), the effect of $T^{\ast}>0$ 
is to split the $SDW$ to $CDW$ transition at $U=2V$ into two parts substituting
a metallic phase along the $U=2V$ line by the LRO dimerized phase for 
$|U-2V|< 8 T^{\ast}/ \pi $.

{\bf Sector ${\bf C}_{{\bf 2}}$}: $U <- |2V+ 8 T^{\ast}/ \pi| $ and $V<0$.
Here, a gap exists in the spin excitation spectrum, the spin field is ordered 
with $\langle \varphi_{s} \rangle = 0$. The charge excitation spectrum is 
gapless and the fixed-point value of the parameter $K_{c} < 1$. The $SS$ 
instability is the dominating one. At $U=2V+8T^{\ast}/\pi$ the spin gap closes.
Triplet superconductivity is no longer suppressed, and $K_{s}=1$, $K_{c}>1$.
 
{\bf Sector D}: $2V+8T^{\ast}/\pi<U<-2V-8T^{\ast}/\pi$. In this sector the 
system shows the properties of the LL metal with dominating 
superconducting istabilities
%%%%%%%%%%%%%%%%%%%%%%%%%%%%%%%%%%%%%%%%%%%
\begin{eqnarray}\label{SS1}
\langle \Delta_{\small SS}(x)\Delta_{\small SS}(x')\rangle 
&=&\langle \Delta_{\small TS}(x)\Delta_{\small
TS}(x')\rangle \nonumber\\
&\sim & \left| x - x' \right|^{-1-1/K_{c}}.
\end{eqnarray}
%%%%%%%%%%%%%%%%%%%%%%%%%%%%%%%%%%%%%%%%%%

Finally, we analyze the case $T^{\ast}<0$ (see Fig. 3b). The phase diagram once
again consists of five different sectors:

Sectors {\bf A}, ${\bf C}_{{\bf 1}}$, ${\bf C}_{{\bf 2}}$, and {\bf D} are 
identical to the corresponding sectors for the $T^{\ast} >0$ case. Particular 
is {\bf sector B}: $|U-2V|<8|T^{\ast}|/\pi$ and $V>0$. In this sector the spin 
spectrum is gapless while the charge excitation spectrum is massive and the 
vacuum expectation value of the ordered charge field is 
$\langle\varphi_{c}\rangle=\sqrt{\pi /8K_{c}}$. Using Eqs. 
(\ref{CDWop})-(\ref{b-SDW}) one obtains in this sector an insulating phase 
with dominating $CDW$ and $bd-SDW$ instabilities showing a power-law decay 
at large distances 
%%%%%%%%%%%%%%%%%%%%%%%%%%%%%%%%%%%%%%%%%%
\begin{eqnarray}
\langle \Delta_{\small CDW}(x)\Delta_{\small CDW}(x')\rangle  &=& 
\langle \Delta_{\small bd-SDW}(x) \Delta_{\small bd-SDW}(x')\rangle 
\nonumber\\ 
& \sim & \left| x - x' \right|^{-1}
\end{eqnarray}
%%%%%%%%%%%%%%%%%%%%%%%%%%%%%%%%%%%%%%%%%
with the critical indices governed by the $SU(2)$--spin symmetry of the model. 

\section{Discussion and summary}

In this paper we have studied the one-dimensional extended Hubbard model with 
CH interactions at half-filling. We have demonstrated that the CH interaction 
can lead to bond located ordering in the ground state. Along the line 
$U=2V>0$, for a "repulsive" three-body coupling ($T^{\ast}>0$), the LRO 
dimerized phase corresponding to an enhanced Peierls instability in the system,
is realized. In the case of an "attractive" three-body term ($T^{\ast}<0$) the 
$bd-SDW$ phase corresponding to a bond located staggered magnetization is -- 
together with the $CDW$ -- the most divergent instability in the system. For 
$T^{\ast}\rightarrow 0$ the sector with new phases shrinks to the line $U=2V$ 
and the ground state phase diagram of the extended Hubbard model 
\cite{Emery1,Voit1,EHM} is recovered. 

Although the phase diagram was studied within the continuum-limit approach, 
assuming the bare-values of the coupling constants much less than the 
bandwidth, the phase diagram is strongly controlled by the symmetry of the 
model. This allows to suppose that the features of the phase diagram will 
persist also in the limit $U,V\gg t$, as far as the ground state phase diagram 
of the extended Hubbard model is essentially the same in both limits 
\cite{Emery1,EHM}. 

However, the weak CH regime is not continuously connected to the ground states 
at $|t^{\ast}| = t$ \cite{ES1,ES2,ES3,ES4,ES5,ES6} implying 
the existence of futher $t^{\ast}$ ($T^{\ast}$) driven phase transitions 
\cite{ES3,ES5,SA1,SA3,SA8}. These additional transition arise only if the 
important finite bandwidth effects are included which is beyond the scope of 
our approach. 

For $V=0$ and $U=U_{c}=8|T^{\ast}|/\pi$ the $SDW$ insulator-metal transition 
is in qualitative agreement with results of numerical studies \cite{SA8}. 
However, contrary to the numerical results showing a TS phase for $U<U_{c}$ 
our results indicate a metallic LL phase in this case, with 
{\it identically} divergent density-density and superconducting instabilities 
due to the $SU(2) \otimes SU(2)$ symmetry of the model. Moreover, due to the 
$SU(2)$-spin symmetry of the model, the dynamical generation of a spin gap for 
$U<-U_{c} < 0$, supports only the SS instabilities. 

For $V \neq 0$ we find that the metallic phase shrinks due to the repulsive nn 
coupling up to the line $U=2V+8|T^{\ast}|/\pi$. There is no superconductivity 
for $V>0$ in the weak coupling phase diagram. 

\centerline{{\bf ACKNOWLEDGMENTS}}

G. J. gratefully acknowledges the kind hospitality at the Center for Electronic
Correlations and Magnetism at the University of Augsburg. This work was 
financially supported by the Deutsche Forschungsgemeinschaft.

\end{document}